\documentclass{pasj00}
\begin{document}
\SetRunningHead{M.\,Sawada}{Suzaku X-ray Observations of Tornado}
\Received{2011/04/28}%{yyyy/mm/dd}
\Accepted{2011/5/20}%{yyyy/mm/dd}
%\Published{}%{yyyy/mm/dd}

\title{Suzaku Discovery of Twin Thermal Plasma from the Tornado Nebula}

\author{
Makoto \textsc{Sawada}\altaffilmark{1}, 
Takeshi Go \textsc{Tsuru}\altaffilmark{1}, 
Katsuji \textsc{Koyama}\altaffilmark{1}, 
and Tomoharu \textsc{Oka}\altaffilmark{2}
}
\altaffiltext{1}{Department of Physics, Graduate School of Science, Kyoto University, 
Kitashirakawa Oiwake-cho, Sakyo-ku, Kyoto 606-8502}
\email{sawada@cr.scphys.kyoto-u.ac.jp}
\altaffiltext{2}{Department of Physics, Faculty of Science and Technology, Keio University, Hiyoshi, 
Kohoku-ku, Yokohama 223-8522}
%% `\KeyWords{}' always has to be placed before `\maketitle'.

\maketitle

\begin{abstract}

The Tornado nebula (G\,357.7$-$0.1) is a mysterious radio 
source with bright ``head'' and faint ``tail'' located 
in the direction of the Galactic center (GC) region. 
We here report the discovery of two diffuse X-ray sources at the 
head and tail of the Tornado with the Suzaku satellite.  
We found emission lines from highly ionized atoms in the two sources. 
The spectra are reproduced by an optically thin thermal plasma with 
a common temperature of 0.6--0.7~keV. 
The interstellar absorption ($N_{\rm H}$) of these sources are the same and are 
slightly larger than that of the GC distance. Since the estimated 
distance using the $N_{\rm H}$ value is consistent with the radio 
observation of the Tornado, these X-ray sources are likely associated with 
the Tornado nebula. The twin-plasma morphology at the both ends of the 
Tornado suggests that the system is a bipolar/outflow source.  

\end{abstract}

\section{Introduction}\label{sec:intro}

G\,357.7$-$0.1 is a bright radio object near the Galactic center (GC) direction and was first catalogued 
by \citet{1960AuJPh..13..676M} as MSH\,17-3{\it9}. 
The extended, non-thermal emission with the spectral index of $\sim -0.6$ (\cite{1973AuJPh..26..379D}, 
\cite{1974IAUS...60..347S}) and the linear polarization of $\lesssim$10\% (e.g. 
\cite{1974AJ.....79..132K}) placed this source as one of the Galactic supernova remnant (SNR).

\begin{figure*}
  \begin{center}
    \FigureFile(160mm,90mm){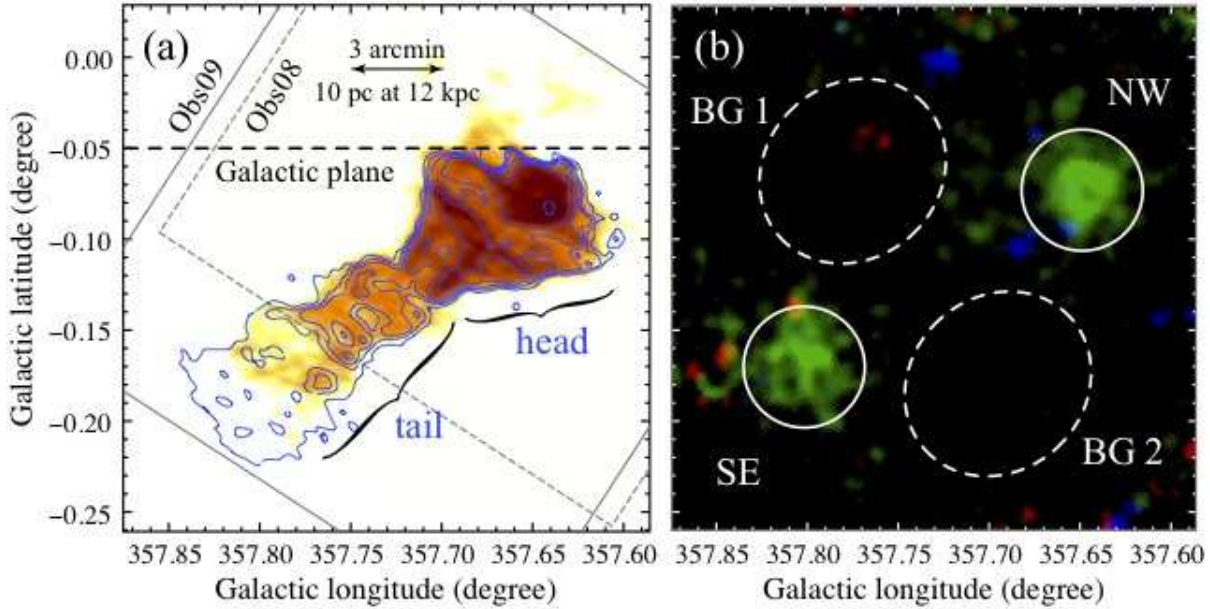}
  \end{center}
  \caption{(a) The 1.4~GHz radio continuum map from the Very Large Array data archives. 
  Overlaid contours represent the radio map with the Australia Telescope Compact Array 
  (ATCA) made by combining 4.79 and 5.84~GHz images (the fainter emission along the 
  tail is enhanced: \cite{1994ApJ...432L..39S}). The XIS fields 
  of view are indicated with the solid and dashed rectangles. 
  (b) Smoothed broad-band XIS images of the Tornado nebula: soft (0.5--1.5~keV in red), 
  medium (1.5--3.0~keV in green), and hard (3.0--7.0~keV in blue) bands are displayed. 
  The data from the three CCDs are merged, NXB are subtracted, and vignetting are corrected. 
  The two fields are combined with normalization of the exposure times. 
  Source and background extraction regions are shown with the solid and dashed ellipses, 
  respectively.}\label{fig:xisimg}
\end{figure*}

It has an elongated shape with $\sim$\timeform{10'} length and has the brightest
peak offset to one side, which is called the ``head'', while fainter emissions are elongated 
to the other side, called the ``tail''. 
Unusual structure was found with the higher resolution observations.
The remarkable feature is  the  several co-axial filamentary arcs 
swirling around its major axis, and hence nicknamed as the ``Tornado nebula'' 
(figure~\ref{fig:xisimg}a: \cite{1985Natur.313..113S}, \cite{1985Natur.313..115B}, \cite{1985Natur.313..118H}). 
\citet{1994ApJ...432L..39S} corrected the large-scale brightness gradient of the radio map, and 
found a bipolar radio structure along the major axis.

The distance to the Tornado is constrained to be $> 6$~kpc from H\emissiontype{I} absorption 
measurements (\cite{1972ApJS...24...49R}). Recently, OH maser at 1720 MHz,
supporting evidence for  the shocked molecular clouds, 
is detected at the head (\cite{1996AJ....111.1651F}, \cite{1999ApJ...527..172Y}, 
\cite{2004MNRAS.354..393L}).  From the velocity of the OH maser 
($-12.4$~km~s$^{-1}$), the distance is estimated to be 11.8~kpc (\cite{1996AJ....111.1651F}). 
The H$_2$ emission lines are also detected toward the head of the Tornado (\cite{2004MNRAS.354..393L}). 

The origin of the Tornado nebula has been controversial due to its peculiar appearance. 
Some scenarios are  based on the head-tail structure; the center of activity is 
located at the brightest head and the tail is the periphery structure. 
The source may be either a radio galaxy with one-sided jet (\cite{1980A&A....90..269W}, 
\cite{1980ARA&A..18..165M}), a Galactic nebula powered by a pulsar, or a Galactic 
binary source located at the head (\cite{1985Natur.313..113S}, \cite{1985Natur.313..115B}, 
\cite{1989ApJ...346..860S}).

The other scenarios are along the lines of bipolar-like structure: an extragalactic double-lobed source 
(\cite{1989PASAu...8..184C}), 
an exotic remnant by an equatorial supernova of a rotating massive star (\cite{1985Natur.313..113S}), 
or precessing jets of binary accretion system (\cite{1985Natur.313..113S}, \cite{1987A&A...171..205M}, 
\cite{1989PASAu...8..184C}). 
The problem of these scenarios is that no compact source has been found 
near at the center of the Tornado.

\begin{table*}
  \caption{Log of Suzaku observations of the Tornado}\label{tab:obslog}
  \begin{center}
    \begin{tabular}{cccccc}
      \hline \hline
      Sequence no. & \multicolumn{2}{c}{Aim point} & Start date & Effective & Field name \\
       & $\alpha$ (J2000.0) & $\delta$ (J2000.0) & & exposure & \\ \hline
      503015010 & \timeform{17h40m07s} & \timeform{-30D57'45''} & 2008/9/19 & 56.8~ks & obs08 \\
      504036010 & \timeform{17h40m31s} & \timeform{-30D56'56''} & 2009/8/29 & 125~ks & obs09 \\
      \hline
    \end{tabular}
  \end{center}
\end{table*}

In the X-ray band, ASCA found a hint of weak X-rays from the Tornado (\cite{2003ApJ...585..319Y}).
Then \citet{2003ApJ...594L..35G} discovered a diffuse X-ray source in the head with Chandra. 
From the centrally filled X-rays, the authors proposed that the Tornado is a Galactic 
mixed-morphology SNR (\cite{1998ApJ...503L.167R}). However the nature of the X-rays, whether 
thermal or non-thermal, was not clear due to the limited photon statistics.

In order to give further constraint on the origin from the high energy phenomena,
we conducted deep X-ray observations on the Tornado nebula 
with the X-ray Imaging Spectrometer (XIS: \cite{2007PASJ...59S..23K}) onboard Suzaku 
(\cite{2007PASJ...59S...1M}). We present the discovery of twin thermal plasmas 
associated with the head and tail of the Tornado. 
We investigate possible interaction of the Tornado with molecular clouds by 
referring the data from the Nobeyama Radio Observatory (NRO) 45-m telescope. 
Based on these new results, we discuss the nature of the X-ray source. 

Throughout this paper, east and west indicate  for the direction in the Galactic longitude, 
and north and south for that in the Galactic latitude. Statistical uncertainties are represented 
by the 90\% confidence intervals.

\section{Observations and Data Reduction}

The Tornado nebula was observed twice. The first observation was centered at the 
head region,  while the second covered the whole structure of the Tornado. The observation 
log is given in table~\ref{tab:obslog}.

The XIS consists of four CCD cameras each placed at the focal planes of four X-Ray 
Telescopes (XRTs: \cite{2007PASJ...59S...9S}). Three sensors employ Front-Illuminated 
(FI) CCDs (XIS\,0, 2, and 3) while 
the other employs a Back-Illuminated (BI) CCD (XIS\,1). XIS\,2 has not been functional since 
an anomaly in 2006 November. The XIS is equipped with spaced-row charge injection technique 
(SCI: \cite{2004SPIE.5501..111B}) to restore the radiation-induced degradation of the energy gain 
and resolution. The field of view (FOV) of XIS combined with XRT covers an \timeform{18'} 
$\times$ \timeform{18'} region with a pixel scale of \timeform{1''}~pixel$^{-1}$. 
An angular resolution of \timeform{1.9'}--\timeform{2.3'} in the half-power 
diameter (HPD) is almost independent of photon energies and off-axis angles 
within $\sim$ 10\%. The total effective area of operational XIS\,0, 1, and 3 
with three XRTs is 430~cm$^2$ at 8~keV. Due to the low orbital altitude 
of Suzaku at $\sim$ 550~km, the XIS achieves a low and stable background environment 
suitable for studying faint diffuse sources. 

In the present observations, the XIS was operated in the normal clocking 
mode with the SCI technique. The data reduction was made from the pipeline
processing version 2.4. We updated the gain correction method to be optimized for 
the SCI (\cite{2009PASJ...61S...9U}) by using the makepi files version 20090915 
provided by the XIS team. 
The systematic uncertainty in the energy scale is $\lesssim$ 10~eV at 
5.9~keV. We removed events during the South Atlantic Anomaly passages 
and Earth night-time and day-time elevation angles below \timeform{5D} 
and \timeform{10D}, respectively. We also removed hot and flickering pixels. 
We reduced the data using the software packages HEADAS version 6.9.

\section{Analysis and Results}

In the spectral analysis, 
we used the Xspec (\cite{1996ASPC..101...17A}) version 12.5.1. 
The non--X-ray background (NXB) data were generated by xisnxbgen (\cite{2008PASJ...60S..11T}). 
The redistribution matrix function is generated by xisrmfgen and the auxiliary 
response function is made by a ray-tracing simulator xissimarfgen (\cite{2007PASJ...59S.113I}).

\subsection{X-ray Images}\label{sec:image}

The NXB subtracted data were used to make the three-band X-ray images 
in the soft (0.5--1.5 keV), medium (1.5--3.0 keV), and hard (3.0--7.0 keV) bands.
Figure~\ref{fig:xisimg}b shows the results after the vignetting correction.
We see two diffuse emissions only in the medium band (1.5--3.0 keV).
We hereafter call these sources as ``NW'' and ``SE''. 
The apparent sizes of NW and SE are $\sim$\timeform{4'}, 
which are larger than the HPD of the Suzaku XRT ($\sim$\timeform{2'}). 
Thus these X-ray sources are not single point sources, but diffuse or ensemble
of multiple point sources. We then searched for X-ray point sources in the source areas 
from the Chandra Catalog (\cite{2010ApJS..189...37E}), and found only one Chandra point source 
in the SE area. The flux is only $0.7$\% of the total flux (after the background subtraction; 
see \S~\ref{sec:bg} and \S~\ref{sec:nwse}) in the SE area 
in the 1.2--7.0~keV band.  We further made the X-ray light-curve from NW and SE areas, 
and found no time variability of these sources.  Thus NW and SE must be   extended sources. 

\subsection{Background Selection}\label{sec:bg}

\begin{figure}[!t]
  \begin{center}
    \FigureFile(80mm,180mm){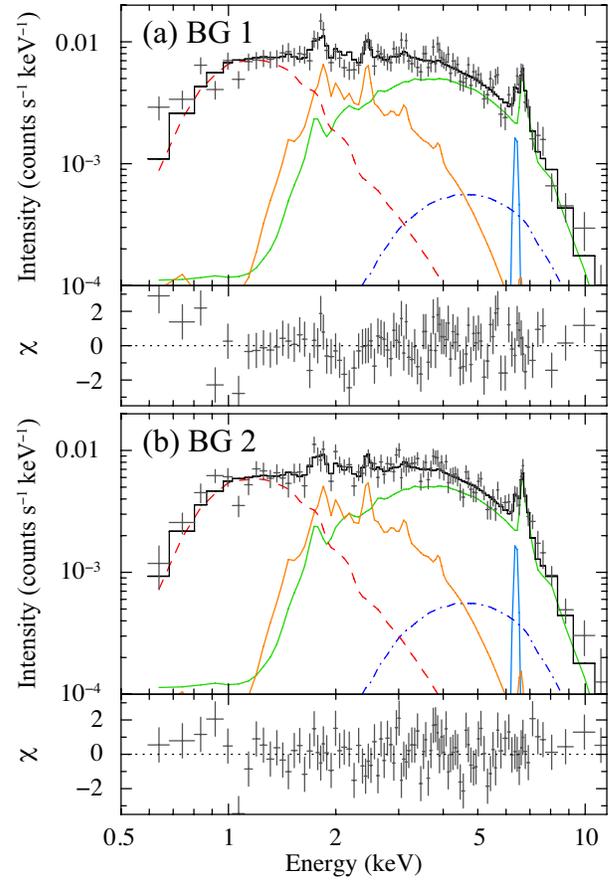}
  \end{center}
  \caption{ (a) The NXB-subtracted spectra (gray crosses) and the best-fit models (black histograms) of  BG\,1. The difference of effective area between the two is corrected and only the results of FIs are shown for simplicity.
  The solid, dashed, and dash-dotted curves represent contributions from the GDXE, the 
  foreground emission, and the CXB, respectively.
(b) Same as panel (a), but for BG\,2.}\label{fig:bgspec}
\end{figure}

The background X-rays in the source (NW and SE) areas are mainly due to the Galactic 
diffuse X-ray emission (GDXE),  because the two sources are near the GC and the Galactic plane. 
The flux (surface brightness) of the GDXE smoothly 
decreases with increasing distances from the GC and the plane, while the spectrum shape 
has no significant position-to-position variation (e.g., \cite{2011PASJ...??S..??U}). 
In order to investigate the flux and spectrum variation near 
the source fields, we select two background regions from the nearby blank sky  (BG\,1 and BG\,2: 
broken ellipses in figure~\ref{fig:xisimg}). Figure~\ref{fig:bgspec} shows the NXB-subtracted spectra of 
BG\,1 and BG\,2.  We see that the over-all spectral structures and fluxes are very similar to each other. 

For the quantitative study, we conducted spectral fittings with the following process. 
The GDXE is known to exhibit strong K-shell lines from highly ionized silicon (Si), sulfur (S), 
and iron (Fe) (\cite{1993ApJ...404..620Y}). 
In fact, the XIS spectra of BG\,1 and BG\,2 clearly exhibit the highly ionized Si, S, and Fe K-shell lines. 
The co-existence of these lines indicates that the spectra require at least two temperature plasmas. 

The previous studies on the GDXE near the GC direction (e.g., \cite{2004ApJ...613..326M}, 
\cite{2009PASJ...61..751R}) showed that the GDXE spectra can be fitted with two plasmas in collisional 
ionization equilibrium (CIE): a low-temperature ($kT\sim1$~keV) CIE for the Si and S lines, and a 
high-temperature ($kT \sim6$~keV) CIE for the Fe line. 
We therefore apply a two-temperature CIE plasma model by using the Apec code (\cite{2001ApJ...556L..91S}). 
We also included a Gaussian in the GDXE as a K$\alpha$ emission line at 6.40~keV from neutral irons 
(Fe\emissiontype{I}), 
which is also found in the GDXE (e.g., \cite{2007PASJ...59S.245K}). 
These three components are attenuated by interstellar extinction (Wabs: \cite{1983ApJ...270..119M}) toward 
the GC distance. Then, the GDXE is written as: 

\begin{eqnarray}
{\rm Wabs}\,1 &\times& ({\rm Apec}\,1+{\rm Apec}\,2+{\rm Gaussian}). 
\end{eqnarray}

The temperatures ($kT$), chemical abundance 
of metals ($Z$) relative to the solar values (\cite{1989GeCoA..53..197A}), 
the volume emission measures of the plasmas, the surface brightness of Fe\emissiontype{I} 
K$\alpha$ line ($S_{\rm 6.4}$), and the hydrogen-equivalent column density of the absorbing 
materials ($N_{\rm{H}}$) were free parameters. 

\begin{table}[!hb]
  \begin{center}
   \caption{Best-fit parameters of the background emission.}\label{tab:bgpar}
   \begin{tabular}{lcc}
      \hline
      Parameter & BG\,1 & BG\,2 \\ \hline
      \multicolumn{3}{c}{--- Low temperature plasma ---}\\
      $kT$ (keV) & 0.90$^{+0.17}_{-0.19}$ & 0.96$^{+0.22}_{-0.17}$ \\
      Norm\footnotemark[$*$] & 8.7$^{+9.6}_{-3.4}$ & 6.4$^{+3.6}_{-2.6}$ \\
       \multicolumn{3}{c}{--- High temperature plasma ---} \\
      $kT$ (keV) & 6.5$^{+1.3}_{-1.3}$ & 6.5$^{+1.2}_{-0.7}$ \\
      Norm\footnotemark[$*$] & 2.3$^{+0.6}_{-0.4}$ & 2.4$^{+0.4}_{-0.3}$ \\
       \multicolumn{3}{c}{--- Common Parameters ---}\\
      $S_{\rm 6.4}$\footnotemark[$\dagger$] & $3.8^{+1.8}_{-1.7}$ & $3.8^{+1.6}_{-1.7}$\\
      $Z$ ($Z_{\odot}$) & 0.53$^{+0.16}_{-0.14}$ & 0.58$^{+0.15}_{-0.13}$ \\
      $N_{\rm{H}}$ ($10^{22}$~cm$^{-2}$) & 5.2$^{+0.8}_{-0.6}$ & 5.3$^{+0.6}_{-0.5}$ \\
      Foreground Norm\footnotemark[$*$] & 1.1$^{+0.1}_{-0.1}$ & 0.91$^{+0.06}_{-0.06}$ \\ \hline      
      $\chi^2$/d.o.f. & 201/171 & 190/171 \\ \hline
      \multicolumn{3}{@{}l@{}}{\hbox to 0pt{\parbox{70mm}{\footnotesize
       \par\noindent
       \footnotemark[$*$] The normalized volume emission measure ($VEM$) represented as 
       $10^{-7}\times \frac{VEM}{4\pi d^2 \cdot A}$~cm$^{-5}$~arcmin$^{-2}$, where $d$ and $A$ are 
       the distance and the solid angle of plasma, respectively. 
       \par\noindent
       \footnotemark[$\dagger$] The average surface brightness in the unit of $10^{-8}$~photons~s$^{-1}$~cm$^{-2}$~arcmin$^{-2}$. 
     }\hss}}
   \end{tabular}
  \end{center}
\end{table}

As is seen in figure 2, the spectra of BG\,1 and BG\,2 exhibit significant emission in the soft band 
(0.5--1.5~keV). If these soft X-rays come from the GDXE, the X-rays should be heavily 
attenuated by interstellar extinction toward the GC distance. Therefore the soft X-rays are mainly 
foreground emissions. The soft X-rays are also observed near the GC region ($|l|<$\timeform{1D}) 
by \citet{2009PASJ...61..751R}.  They found that the soft X-ray emission is almost uniform within the 
central one degree on the plane. 
They applied the model below: 

\begin{equation}
{\rm Wabs}\,2\times {\rm Apec}\,3.
\end{equation}
\noindent

Since the absolute Galactic latitude of the Tornado fields are close to that of the region of 
\citet{2009PASJ...61..751R},  we can reasonably assume that the typical 
parameters for the soft X-ray spectrum are same as those of \citet{2009PASJ...61..751R}: 
the plasma temperature of 0.9~keV, the chemical abundance\footnote{Since the true origin 
of the foreground emission is unknown at present, we applied a phenomenological model, 
which gives a reasonable fit to the data. Therefore apparent low abundances in this model 
should not be taken seriously.} of $5\times10^{-3}$~$Z_{\odot}$, 
and the hydrogen-equivalent absorption column of 
$2.0\times10^{21}$~cm$^{-2}$.  Thus for the foreground soft X-ray spectrum, only the 
normalization was a free parameter.

The other background component, the Cosmic X-ray background (CXB), is isotropic and 
we adopted the empirical form, an absorbed power-law:

\begin{equation}
{\rm Wabs}\,3 \times C \times (E/{\rm keV})^{-\Gamma},
\end{equation}
\noindent
where $C$ is the normalization constant at 1~keV and $\Gamma$ is the photon index of the power-law.  
These are fixed 
 to be  $C=7.4\times10^{-7}$~photons~s$^{-1}$~cm$^{-2}$~arcmin$^{-2}$ 
and $\Gamma=1.486$ according to \citet{2002PASJ...54..327K}. 
Since the CXB should be absorbed by the interstellar materials throughout the whole 
Galaxy,  we assumed the hydrogen-equivalent absorption column toward the CXB to be twice 
of that for the GDXE ($N_{\rm H}$).
 
In the model fitting, all the free parameters are allowed to be independent 
between the two regions, BG\,1 and BG\,2. The model fit was acceptable for both the spectra.
The results are shown in figure~\ref{fig:bgspec} and the best-fit parameters are listed in 
table~\ref{tab:bgpar}.  

We found that the best-fit parameters for the GDXE components are all consistent between BG\,1 and 
BG\,2 within the statistical errors, and hence can assume that the background spectrum and flux 
are essentially constant near at the source areas.
We therefore merged BG\,1 and BG\,2 to increase statistics. The best-fit $N_{\rm H}$ for the merged spectrum is  $(5.3\pm0.4)\times 10^{22}$~cm$^{-2}$.
 
\begin{figure}[!h]
  \begin{center}
    \FigureFile(80mm,180mm){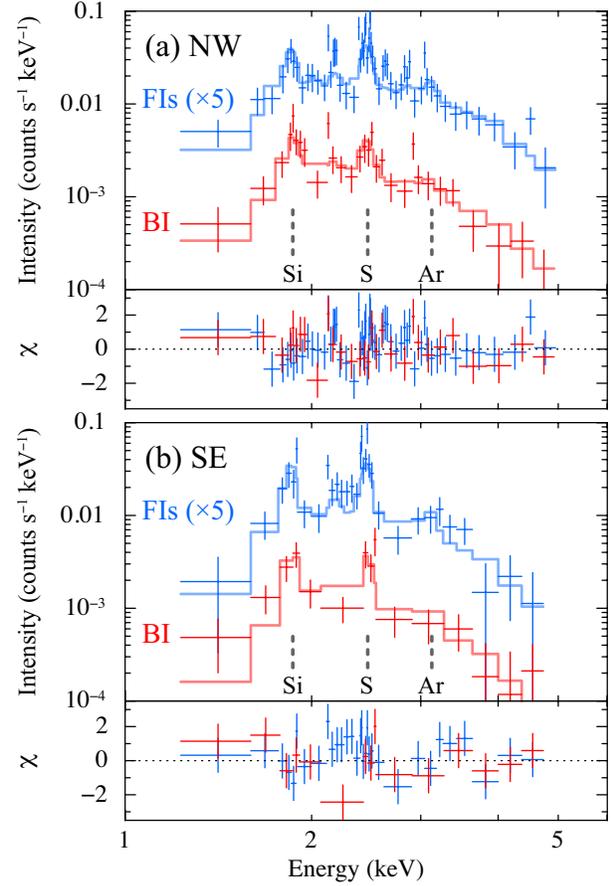}
  \end{center}
  \caption{(a) The background-subtracted spectra for NW. The FI and BI spectra are separately 
  given by blue and red crosses, respectively. For visibility, the FI spectrum is multiplied by~5. 
  The blue and red histograms are the best-fit model of FI and BI spectra, respectively.
 (b) Same as panel (a) but for SE.}\label{fig:spec} 
\end{figure}

\begin{table}[!hb]
  \begin{center}
   \caption{Best-fit parameters of the diffuse sources.}\label{tab:specpar}
   \begin{tabular}{lcc}
      \hline
      Parameter & NW & SE \\ \hline
      $N_{\rm{H}}$ (10$^{22}$ cm$^{-2}$) & 6.6$^{+1.1}_{-0.8}$ & 7.4$^{+1.7}_{-1.4}$\\
      $kT$ (keV) & 0.73$^{+0.15}_{-0.15}$ & 0.59$^{+0.18}_{-0.15}$ \\
      $Z$ ($Z_{\odot}$)\footnotemark[$*$] & 0.89$^{+0.37}_{-0.28}$ & 1.7$^{+1.3}_{-0.7}$\\ 
      $S_{\rm{X}}$\footnotemark[$\dagger$] & 2.8$^{+0.2}_{-0.2}$ & 2.4$^{+0.2}_{-0.2}$ \\ 
      $L_{\rm{X}}$\footnotemark[$\ddagger$] & 2.4$^{+0.1}_{-0.1}$ & 3.6$^{+0.3}_{-0.3}$ \\ \hline
      $\chi^2$/d.o.f. & 85/82 & 49/41\\ \hline
       \multicolumn{3}{@{}l@{}}{\hbox to 0pt{\parbox{75mm}{\footnotesize
      \par\noindent
      \footnotemark[$*$] The chemical abundance of Si, S, and Ar relative to the solar values. The abundances of the other elements are fixed at the solar values.
      \par\noindent
       \footnotemark[$\dagger$] The average surface brightness in the unit of $10^{-6}$~photons~s$^{-1}$~cm$^{-2}$~arcmin$^{-2}$ in the 0.5--7.0~keV band.  
       \par\noindent
       \footnotemark[$\ddagger$] The luminosity in the unit of $10^{35}$~erg~s$^{-1}$ in the 0.5--7.0~keV band. The absorption is corrected and the distances are assumed to be 12.0~kpc (see \S~\ref{sec:nature}).
     }\hss}}
   \end{tabular}
  \end{center}
\end{table}

\subsection{Spectra of NW and SE}\label{sec:nwse}

We extracted the spectra from the NW and SE regions (solid circles in figure 1b), 
then subtracted the NXB and merged-background spectra.
Figure~\ref{fig:spec} shows the background-subtracted spectra of NW and SE. 
We clearly see several emission lines from highly ionized atoms,  Si, S and argon (Ar). 
We therefore fitted the spectra with an optically thin thermal 
plasma model in CIE (Apec) absorbed by the interstellar medium (Wabs). 
Free parameters were the chemical abundances of Si, S, and Ar, the temperatures
and normalization of the CIE plasma, and absorption ($N_{\rm H}$).
The abundances of the  elements other than Si, S, and Ar 
were fixed to the solar values. All the abundances in the 
absorbing materials were also fixed to be solar. 
With this model, we obtained satisfactory fits both for NW and SE (figure~\ref{fig:spec}). 
The best-fit parameters are given in table~\ref{tab:specpar}.

\section{Discussion}

\subsection{Nature of the X-ray Sources and Association to the Tornado}\label{sec:nature}

We find that the absorption column densities ($N_{\rm H}$) for the NW and SE sources are very similar to
each other (table 3).  Since the $N_{\rm H}$ values for the GDXE near the source region are
almost constant at $(5.3\pm0.4)\times 10^{22}$~cm$^{-2}$,
  the same $N_{\rm H}$ values for NW and SE supports that the two sources 
are located at nearly the same distances. 

\begin{figure}[!h]
  \begin{center}
    \FigureFile(80mm,90mm){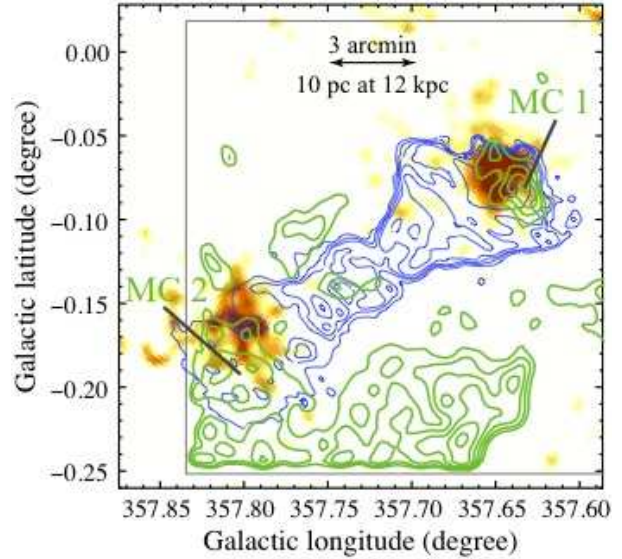}
  \end{center}
  \caption{Multi-wavelength image of the Tornado nebula: the medium band (1.5--3.0~keV) image overlaid with contours of the radio continuum emission 
  with the ATCA (blue) and CO $J$=1--0 ($V_{\rm{LSR}}=-11\pm1$~km~s$^{-1}$) with the NRO 45~m telescope (green). 
The FOV of the NRO observation is shown with the solid rectangle.}
\label{fig:multiimg}
\end{figure}

Assuming that the distance of the GDXE is 8~kpc and that the interstellar gas 
density along the line-of-sight is constant, the distances of NW and SE can be estimated by the 
$N_{\rm H}$ ratio relative to the background emission of ($5.3\pm0.4)\times 10^{22}$~cm$^{-2}$ as 
$10.0^{+1.8}_{-1.4}$ and $11.2^{+2.7}_{-2.3}$~kpc, respectively.
These distances are almost consistent with  the radio distance of the Tornado ($\sim$12~kpc; \cite{1996AJ....111.1651F}).
In the projected image, the radio emission of the Tornado appears to be associated with the two X-ray structures at the both ends (see figure~\ref{fig:multiimg}).  
Thus we argue that the two X-ray sources are physically connected to the Tornado nebula. 

The two diffuse sources are found to exhibit optically thin thermal nature. 
In fact, the X-ray emission from the sources are described with CIE plasmas 
of nearly the same temperatures (0.6--0.7~keV). The abundances are consistent with  the solar values. 
This indicates that the X-ray emitting materials mainly consist of interstellar gas.
If the distance of the Tornado is 12~kpc, the X-ray source size of \timeform{4'} corresponds to 14~pc.  Assuming a spherical shape with uniform density, we derive the physical parameters of the NW and SE plasmas, electron density
($n_{\rm e}$), mass ($M$) and total thermal energy ($E_{\rm th}$),
as are listed in table~\ref{tab:plasmapar}.

We found no hint of non-thermal emission from the NW and SE sources.
Assuming the photon index of 2, we obtained the 3-$\sigma$ upper limits 
of the luminosities of non-thermal X-rays to be $1.1$\% and $2.0$\% of those of the 
thermal plasmas in the 0.5--7.0~keV band for NW and SE, respectively. 

\begin{table}[!ht]
  \begin{center}
   \caption{Physical parameters of the X-ray emitting plasma.}\label{tab:plasmapar}
   \begin{tabular}{lcc}
      \hline
      Parameter & NW & SE \\ \hline
     $VEM$ ($10^{58}~$cm$^{-3}$)\footnotemark[$*$]& 1.0 & 1.5 \\
      $n_{\rm e}$ (cm$^{-3}$) & 0.49 & 0.59 \\
      $M$ ($M_{\odot}$) & 23 & 29 \\
      $E_{\rm th}$ ($10^{49}$~erg) & 8.3 & 8.2 \\ \hline
       \multicolumn{3}{@{}l@{}}{\hbox to 0pt{\parbox{70mm}{\footnotesize
       \par\noindent
       \footnotemark[$*$] These values are taken from the best-fit model in table 3 at the distance of 12~kpc.}\hss}}
   \end{tabular}
  \end{center}
\end{table}

\subsection{Bipolar Structure of the Tornado Nebula}\label{sec:bipolar} 

The two X-ray plasmas are located at the both ends (head and tail) of the symmetric 
(bipolar-like) radio structure found by \citet{1994ApJ...432L..39S}.
Since the properties of the X-ray plasmas are almost identical, 
these would be the twins formed by a common process. 
A plausible scenario is that the X-ray plasmas have been generated by the 
termination shocks  at the intersecting molecular clouds. 
In fact the shocked molecular cloud associated with the Tornado has been found  near NW (\cite{1996AJ....111.1651F}).  To confirm the cloud interactions,  
we further search for the molecular structure near at the  NW and 
SE regions using the data of CO $J=$1--0 taken with the NRO 45-m telescope 
(private communication with T. Oka). The results in the velocity range of $11\pm1$~km~s$^{-1}$, 
close to the velocity of the OH maser near at NW, are shown in figure~\ref{fig:multiimg}.
We clearly see CO counterparts near NW (MC\,1) and SE (MC\,2).  

%%%%%%%%%%%%%%%%%
As we described in \S~\ref{sec:intro} (Introduction), many ideas for the origin of the Tornado have been 
proposed. The present result that the Tornado has bipolar X-rays at the
head and tail excludes most of these scenarios. 
In the following sections (\S~\ref{sec:jet} and \S~\ref{sec:remnant}), we discuss two plausible origins.

\subsection{Bipolar Jet from a Binary Accretion System}\label{sec:jet}

One possibility is that the Tornado is made by bipolar jets from a binary accretion system.
The filamentary radio arcs swirling around the major axis would be due to the precession of 
jets as is found in SS\,433. The twin X-ray emitting plasmas are made by the shock with molecular clouds.
Bipolar jets have usually a bright central power source, but no X-ray source is found from the center of 
the Tornado nebula. 
We set the 3-$\sigma$ upper limit on the luminosity to be $1.9\times 10^{33}$~erg~s$^{-1}$ 
in the 0.5--10.0~keV band, assuming the absorption column of $N_{\rm H}=7\times10^{22}$~cm$^{-2}$, 
the power-law spectrum with the photon index of 2, and the distance of 12~kpc. 
Chandra has gave more severe constraint of  $\sim1\times10^{33}$~erg~s$^{-1}$ (\cite{2003ApJ...594L..35G}). 
These upper limits are only less than 1\% of  NW and SE luminosities
of $2.4\times 10^{35}$~erg~s$^{-1}$ and $3.6\times 10^{35}$~erg~s$^{-1}$, respectively.
A plausible explanation for the non-detection of the bright central source is that putative
binary went into a quiescent state after forming jet-structures in a past active phase.

The synchrotron cooling timescale (\cite{1996ApJ...459L..13R}) for electrons which emit
$\epsilon_{\rm syn}$~keV photons in a magnetic field $B$~G  is estimated to be, 
\begin{eqnarray}
\tau \approx 2000\left( \frac{B}{10\ \mu{\rm G}} \right)^{-\frac{3}{2}} \left(\frac{\epsilon_{\rm syn}}
{1\ {\rm keV}}\right)^{-\frac{1}{2}}\ {\rm years}.
\end{eqnarray}
Assuming the magnetic field of $100$~$\mu$G (\cite{2008ApJS..177..255L}), 
the cooling timescale for $\ge 1$ keV photons is estimated to be less than $\sim$40--50~years.
The lower energy electrons responsible for the radio synchrotron radiation can survive much longer time. 
The cooling time of the X-ray emitting plasma, which are made by the interaction between the jets and 
the molecular clouds, is also much longer.
Thus the radio Tornado and the twin X-rays at the both ends would be fossil 
radiations of a past activity some  $\sim$40--50~years ago.
Although such object is very rare, we see a hint in the jets of 4U\,1755$-$33. 
The central source is in low luminosity of $3.6\times10^{31}$~erg s$^{-1}$ at present,
but was bright two decades ago (\cite{2003ApJ...586L..71A}, \cite{1996IAUC.6302....2R}).

\subsection{Remnant of an Equatorial Explosion}\label{sec:remnant}

An alternative possibility for the bipolar structure of the Tornado is 
a remnant of an equatorial explosion of a rotating high mass star 
(\cite{1985Natur.313..113S}). 
For example, a hypernova explosion associated with a 
$\gamma$-ray burst may produce a bipolar blast wave (\cite{2004ApJ...613L..17I}). 
If this blast wave hits molecular clouds, then the double X-ray spots will be produced.
This scenario does not necessary require a central source.
However, a hypernova is very energetic, and in our case, 
the explosion energy is less than $10^{51}$ erg s$^{-1}$ (table 4). 
We therefore suspect this possibility is less likely compared 
to the bipolar-jet model of an X-ray binary. 

\section{Summary}

\begin{enumerate}

\item 
We detected two diffuse X-ray sources and obtained the
high quality spectra with the XIS. 
Both the spectra exhibit clear emission lines from highly ionized atoms, 
which establish the optically thin hot plasma nature. The properties 
of the plasmas are almost the same with each other. 

\item 
We derived the distances to the sources by using the 
X-ray absorption and found that both the sources are the Galactic objects located slightly 
beyond the GC distance, at the consistent distance with the radio object, Tornado. 
Thus, together with the association in the projected image,
the two X-ray sources are physically connected to the Tornado nebula. 

\item 
The discovery of the twin thermal plasmas at the head and tail of the Tornado 
confirm the symmetric structure with respect to the center.

\item 
The most plausible origin is a bipolar jet from an accreting binary 
which had been active in the past and has recently entered the quiescent state. 
In this view, the X-ray emission is a fossil 
radiation from shock heated gas made by the past jet activity 
of the central core source. The non-thermal synchrotron X-rays may be
already cooled down below the detection limit. 

\end{enumerate}

\bigskip

The authors thank all of the Suzaku team members for their full support of the Suzaku project. 
We acknowledge the financial support from the Ministry of Education, Culture, Sports, Science and Technology (MEXT) of Japan; the Grant-in-Aid for the Global COE Program ``The Next Generation of Physics, Spun from Universality and Emergence'' and others for Scientific Research B (No. 20340043 and 23340047).　
KK is supported by the Challenging Exploratory Research program. MS is supported by Japan Society for the Promotion of Science (JSPS) Research Fellowship for Young Scientists.

\end{document}